\documentclass[12pt]{iopart}

\usepackage[pdftex]{graphicx} 
\usepackage{subfigure}
\begin{document}
\title[Dihadron Correlations in PbPb]{Dihadron correlations in PbPb collisions at $\sqrt{s_{NN}}$ = 2.76 TeV with CMS}

\author{Jeremy Callner for the CMS Collaboration}

\address{University of Illinois at Chicago, Chicago, IL, USA}
\ead{jcallner@gmail.com}
\begin{abstract}
Measurements of charged dihadron $\Delta\eta$$-$$\Delta\phi$ correlations from the CMS collaboration are presented 
for PbPb collisions at a center-of-mass energy of 2.76 TeV per nucleon pair over a broad range of 
pseudorapidity and the full range of azimuthal angle. 
A significant correlated yield is observed for pairs of particles with small $\Delta\phi$ but 
large longitudinal separation $\Delta\eta$, commonly known as the ``ridge''. The ridge persists up 
to at least $|\Delta\eta|$ = 4 and the dependence of the ridge region shape and yield on
collision centrality and transverse momentum has been measured. 
A Fourier analysis of the long-range two-particle correlation is presented and 
discussed in the context of higher order flow coefficients.
\end{abstract}

The starting point of this dihadron correlations analysis is the two-dimensional (2-D)
per-trigger-particle associated yield as a function of
$\Delta\eta$ and $\Delta\phi$ for all particles falling within a 
given ``trigger'' (\ensuremath{p_T^\mathrm{trig}}) 
and ``associated'' (\ensuremath{p_T^\mathrm{assoc}})
transverse momentum interval.
The CMS silicon tracker~\cite{CMS:2008zzk}, 
with its large pseudorapidity coverage, 
is ideally suited for detailed analyses of both short- and long- range 
charged hadron correlations at the LHC. 
Dihadron correlation results from CMS in the 0--5\% most central PbPb collisions
have recently been reported in Ref.~\cite{ref:HIN-11-001-PAS}, 
which found several non-trivial features including 
a significant broadening of the away side 
($\Delta\phi$$>$ 1) dihadron correlation when compared to pp collisions and
a prominent near side ridge centered at $\Delta\phi$$\approx$0 with 
an associated yield that decreases as a function of increasing transverse 
momentum above 6~\ensuremath{\mathrm{GeV}/c}. 
The current analysis extends the results to 12 centrality classes and
also explores in detail the resulting structure and its possible connection
with collective motion as manifest in a Fourier analysis
of the $\Delta\phi$ associated yield distribution.
  
\begin{figure}[ht]
  \begin{center}
    \includegraphics[width=.8\linewidth]{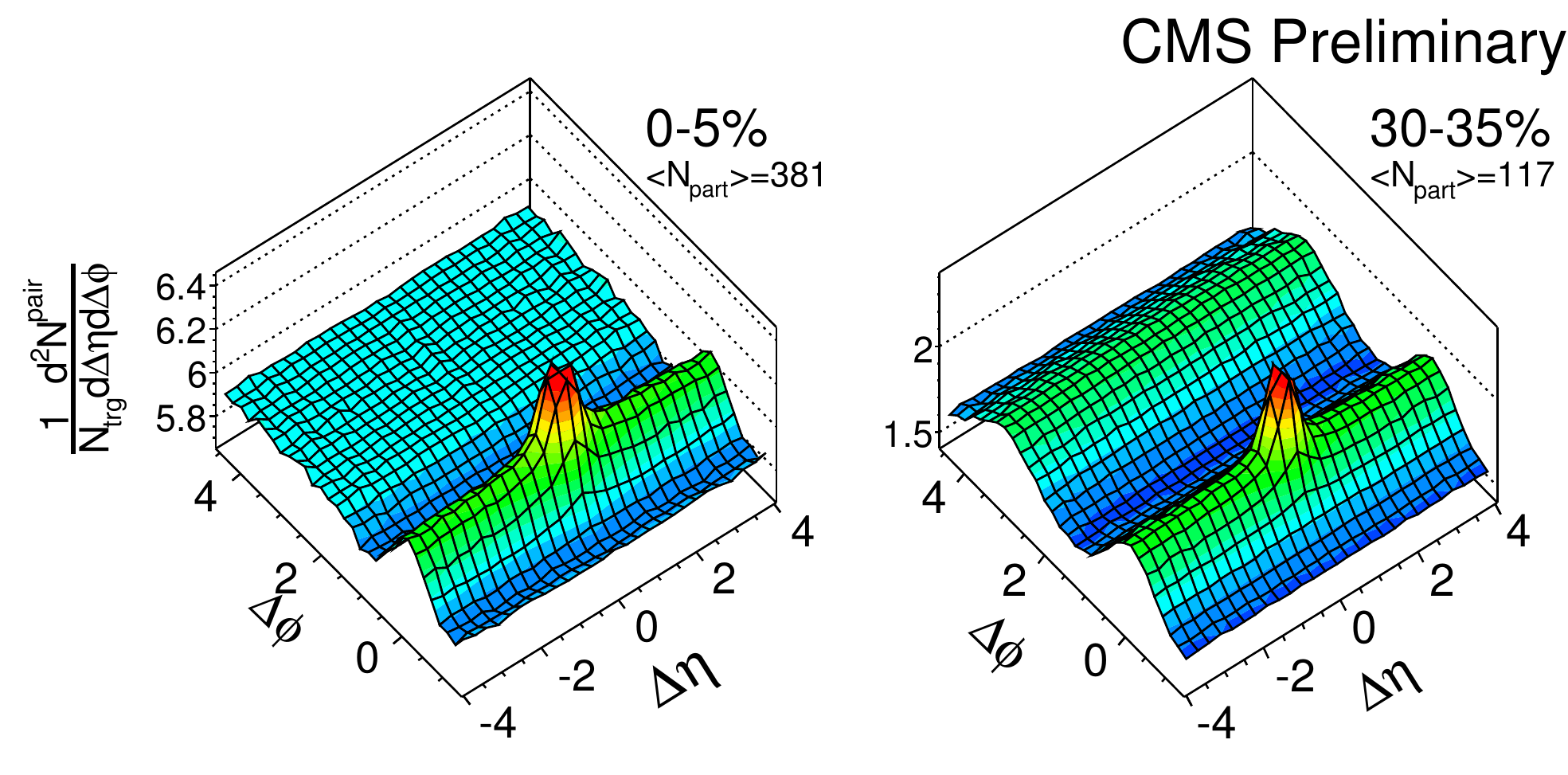}
    \vspace{-0.4cm}
    \caption{ Two-dimensional (2-D) per-trigger-particle associated yield of charged hadrons
         as a function of $\Delta\eta$ and $\Delta\phi$ for 
         $4 $$<$$ \ensuremath{p_T^\mathrm{trig}} $$<$$ 6$~\ensuremath{\mathrm{GeV}/c} and 
         $2 $$<$$ \ensuremath{p_T^\mathrm{assoc}} $$<$$ 4$~\ensuremath{\mathrm{GeV}/c} in 
         central (left) and mid-peripheral (right) PbPb collisions at 
         \ensuremath{\sqrt{s_{_{NN}}}} = 2.76~TeV. 
    }
    \label{fig:fig1}
  \end{center}
\vspace{-0.8cm}
\end{figure}

An example of the resulting 2-D correlations for trigger particles with 
$4$$<$$\ensuremath{p_T^\mathrm{trig}}$$<$$6~\ensuremath{\mathrm{GeV}/c}$ and associated particles 
with $2$$<$$\ensuremath{p_T^\mathrm{assoc}}$$<$$4~\ensuremath{\mathrm{GeV}/c}$ is shown 
in Fig.~\ref{fig:fig1} for two selected centrality classes.
The 2-D associated yield distributions 
reveal a rich structure and evolution 
as a function of centrality.
Of most relevance to the current investigation is 
the ridge yield centered at $\Delta\phi $$\approx$$0$ and
extending to $|\Delta\eta|$$=$$4$.
In mid-peripheral collisions, a pronounced $\cos(2\Delta\phi)$ component
emerges, presumably originating from the elliptic flow effect and
similar in character to what has been observed at 
RHIC energies ~\cite{star:2005ja,phenix:2007aa,Alver:2008gk}.

\begin{figure}[htb]
  \begin{center}
    \subfigure[]{\includegraphics[width=.72\textwidth]{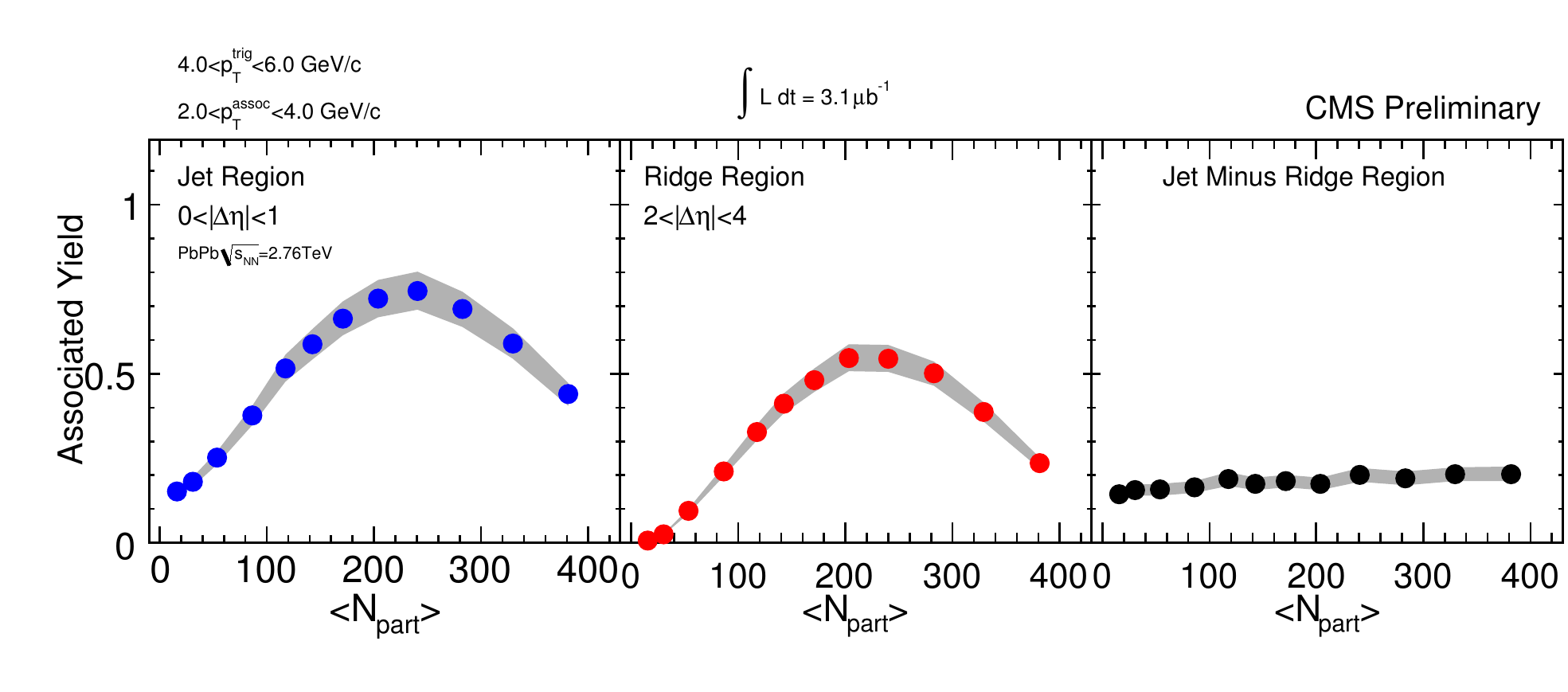}}
    \subfigure[]{\includegraphics[width=.26\textwidth]{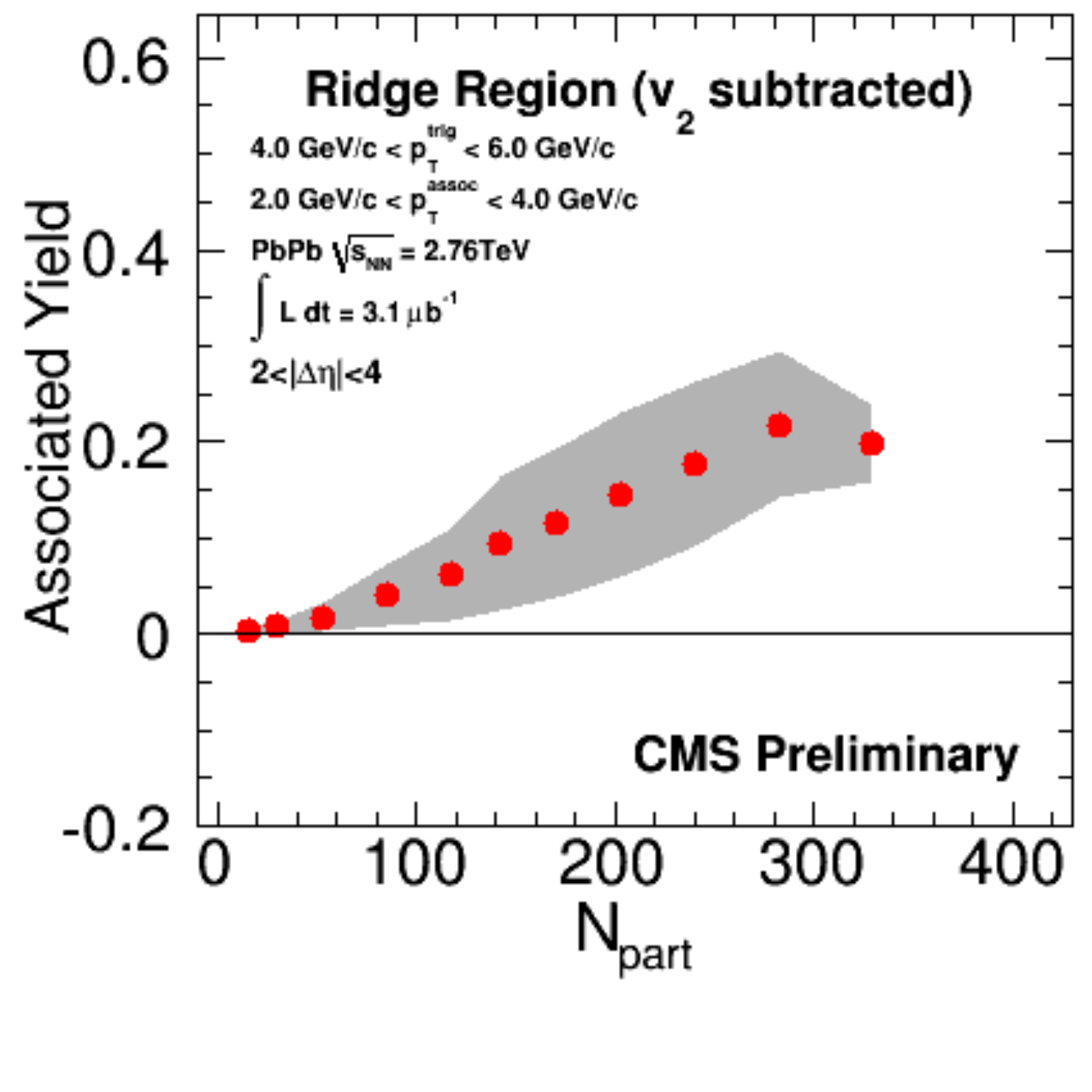}}
    \caption{  
      (a) Integrated near-side
      associated yields for $4 $$<$$ \ensuremath{p_T^\mathrm{trig}} $$<$$ 6~\ensuremath{\mathrm{GeV}/c}$ 
      and $2 $$<$$ \ensuremath{p_T^\mathrm{assoc}} $$<$$ 4~\ensuremath{\mathrm{GeV}/c}$,
      above the minimum level found by the ZYAM procedure,
      for (left) the jet region ($0$$<$$|\Delta\eta|$$<$$1$), 
      (middle) the ridge region ($2$$<$$|\Delta\eta|$$<$$4$), 
      and difference of jet-ridge region (right) as a function 
      of \ensuremath{N_{part}} for PbPb collisions at \ensuremath{\sqrt{s_{_{NN}}}} = 2.76 TeV. 
      (b) Integrated near-side associated yield for the ridge region with the CMS measured
      elliptic flow $v_2$ component subtracted.
      The gray bands correspond to systematic uncertainties \cite{ref:HIN-11-006-PAS}.
    }
    \label{fig:fig2}
  \end{center}
\vspace{-0.8cm}
\end{figure}

To quantitatively examine the features of short- and long- range azimuthal correlations,
one dimensional (1-D) $\Delta\phi$ correlation functions are calculated by averaging 
the 2-D distributions over a selected region in $\Delta\eta$.
The zero-yield-at-minimum (ZYAM) methodology applied to the resulting
1-D $\Delta\phi$ correlation enables the extraction of the integrated associated yields
in the near-side jet and ridge regions around $\Delta\phi$$\approx$$0$, as
outlined in Ref.~\cite{ref:HIN-11-001-PAS}.
The ZYAM methodology is also applied in this analysis for the $\Delta\phi$
distributions, with the results shown in Fig.~\ref{fig:fig2}(a). 
Two distinct regions in $|\Delta\eta|$ are studied independently;
the short-range ``Jet Region'' centered at 
$(\Delta\eta,\Delta\phi) $$\approx$$ (0,0)$ and the
long-range ``Ridge Region'' at larger $|\Delta\eta|$ from 2 to 4 units.
A strong centrality dependence is observed in
the associated yields that comes primarily from the long-range ridge region. By
subtracting off the ridge region yield from the jet region yield, 
the residual jet region associated yield is found to be largely independent of centrality, 
a general feature of the data that is similar to that seen at RHIC~\cite{star:2007pu,star:2010ridge}.

The pronounced $\cos(2\Delta\phi)$ component visible in the 2-D 
correlations has been associated with the 
elliptic flow $v_2$ collective motion, and thus it is desireable
to subtract this component from the dihadron 
correlations in order to study the residual yield \cite{star:2009qa}.
This analysis was performed utilizing the CMS results
for elliptic flow \cite{ref:HIN-10-002-PAS} with additional
details on the subtraction methodology given in Ref.~\cite{ref:HIN-11-006-PAS}.  
The elliptic flow $v_2$ subtracted associated yield for the ridge
region is given in Fig. \ref{fig:fig2}(b).
The most prominent feature of the $v_2$-subtracted 
result is that the remaining ridge region yield is found to be 
negligible for lower \ensuremath{N_{part}} values, and increases roughly 
linearly to an \ensuremath{N_{part}} value of around 300. 
This general feature of the $v_2$-subtracted ridge yield is 
similar to that seen at lower energies \cite{Alver:2009id}, where the associated
yield was found to drop to zero below \ensuremath{N_{part}}$\approx$100.  
This analysis 
indicates that many of the centrality dependent features of the ridge yield 
in PbPb collisions at \ensuremath{\sqrt{s_{_{NN}}}} = 2.76~TeV, for 
our selected trigger and associated momentum ranges, are very similar in
character to what has been observed in lower energy
AuAu collisions at \ensuremath{\sqrt{s_{_{NN}}}} = 0.2~TeV.
   
Motivated by the idea of understanding the 
long-range ridge effect in the context of higher-order hydrodynamic flow 
(e.g. induced by initial geometric fluctuations \cite{Alver:2010gr}) an alternative way of quantifying 
the observed long-range correlations using a Fourier decomposition technique is 
investigated.  Following the methodology of \cite{ref:HIN-11-001-PAS}, 
the $\Delta\phi$ distribution
is decomposed into a Fourier series using the following expression:

\begin{equation}
\label{fourier}
\frac{1}{N_{\rm trig}}\frac{dN^{\rm pair}}{d\Delta\phi} = \frac{N_{\rm assoc}}{2\pi} \left ( 1+\sum\limits_{n=1}^{\infty} 2C^{\rm f}_{n} \cos (n\Delta\phi)\right ),
\end{equation}

\noindent 
where $N_{\rm assoc}$ represents the total number of dihadron pairs per trigger 
particle for a given $|\Delta\eta|$ range and (\ensuremath{p_T^\mathrm{trig}}, 
\ensuremath{p_T^\mathrm{assoc}}) bin. The result of the Fourier analysis 
is that the observed 1-D $\Delta\phi$ projections in the ridge region 
are well described by the first five terms in the series.
If the observed correlation is purely driven by the single-particle azimuthal
anisotropy arising from the hydrodynamic expansion of the medium~\cite{Voloshin:1994mz}, 
the extracted Fourier $C^{\rm f}_{n}$ components would be related to the flow coefficients 
$v_{n}$ via: 

\begin{equation}
\label{eq:factorization}
C^{\rm f}_{\rm n}(\ensuremath{p_T^\mathrm{trig}}, \ensuremath{p_T^\mathrm{assoc}}) = v^{\rm f}_{\rm n}(\ensuremath{p_T^\mathrm{trig}}) \times v^{\rm f}_{\rm n}(\ensuremath{p_T^\mathrm{assoc}}),
\end{equation}

\noindent 
where $v^{\rm f}_{\rm n}(\ensuremath{p_T^\mathrm{trig}})$ 
and $v^{\rm f}_{\rm n}(\ensuremath{p_T^\mathrm{assoc}})$ can be identified
with the flow coefficients for the trigger and associated particles
as extracted from this type of Fourier analysis.
Non-flow effects on the extracted $v^{\rm f}_{\rm n}$ values 
can be minimized by fixing the associated particles at low $p_T$ values
where hydrodynamic flow should dominate. 
Following the assumption in Eq.~\ref{eq:factorization}, the hydrodynamic 
flow $v^{\rm f}_{\rm n}$ harmonics as a function of \ensuremath{p_T^\mathrm{trig}} can be extracted
from the measured $C^{\rm f}_n$ values as follows:

\begin{figure}[thb]
  \begin{center}
    \subfigure[]{\includegraphics[width=0.55\linewidth]{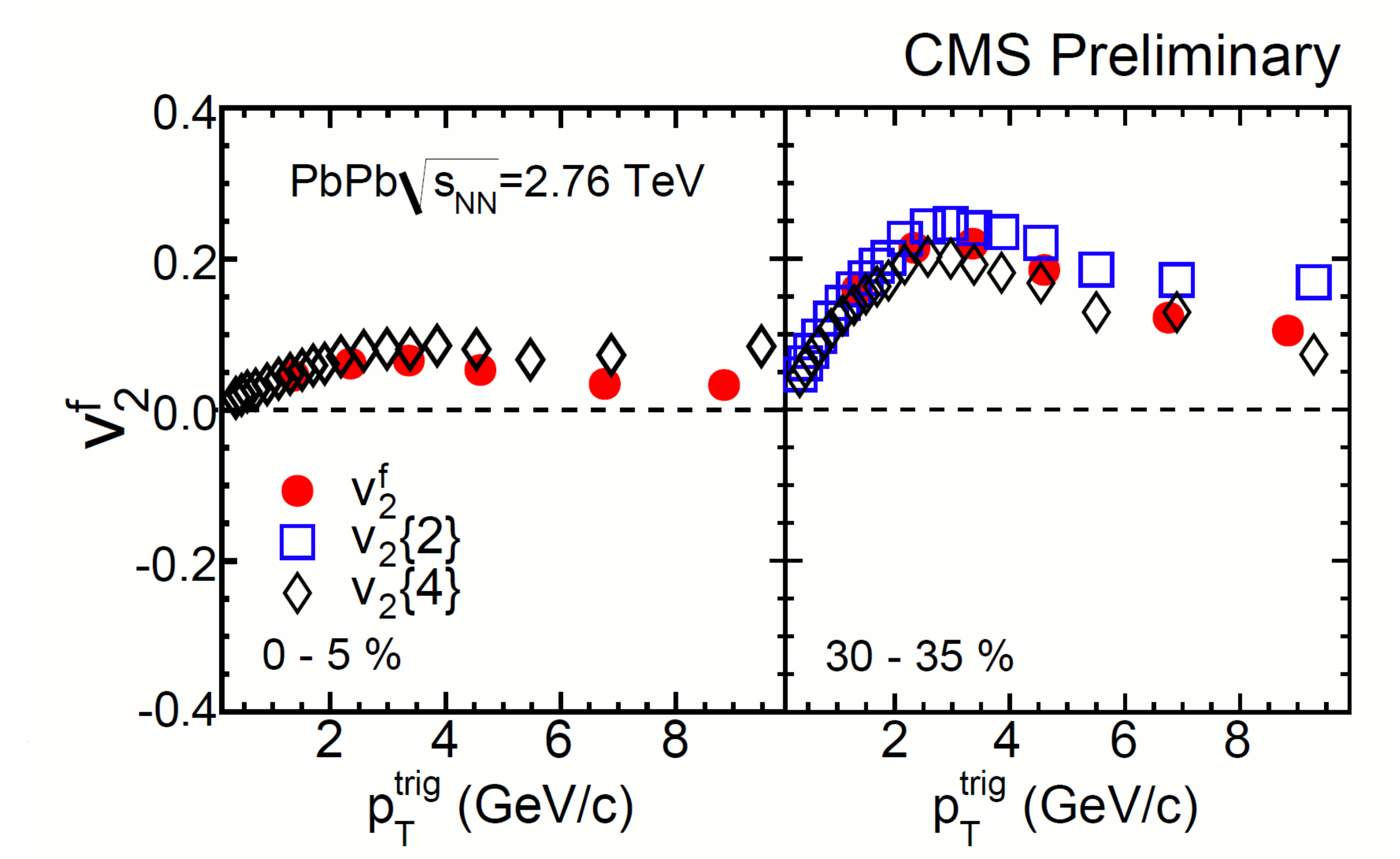}}
    \subfigure[]{\includegraphics[width=.40\linewidth]{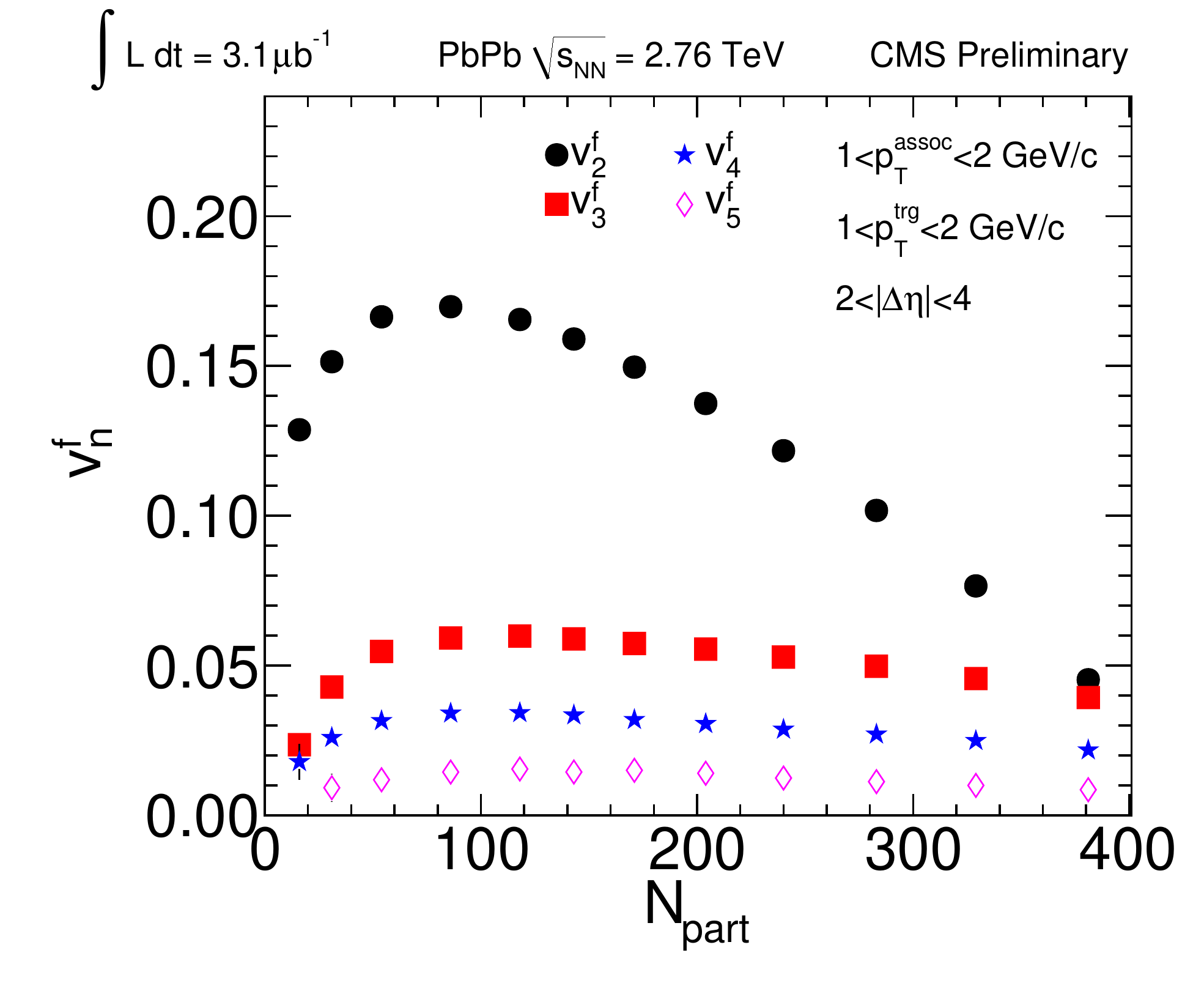}}
    \caption{(a) $v^{\rm f}_{2}$ hydrodynamic harmonic extracted
      as a function of \ensuremath{p_T^\mathrm{trig}} for 
      $1$$<$$\ensuremath{p_T^\mathrm{assoc}}$$<$$2$~GeV/c for central (left) and mid-peripheral (right)
      PbPb collisions at \ensuremath{\sqrt{s_{_{NN}}}} = 2.76~TeV. 
      Comparisons to the two- and four-particle cumulant method of measuring $v_{2}$ 
      (b) The flow harmonics $v^{\rm f}_{2}$, $v^{\rm f}_{3}$, $v^{\rm f}_{4}$ and 
      $v^{\rm f}_{5}$ extracted from long-range ($2$$<$$|\Delta\eta|$$<$$4$) 
      azimuthal dihadron correlations for $1$$<$$\ensuremath{p_T^\mathrm{trig}}$$<$$2$~\ensuremath{\mathrm{GeV}/c} 
      and $1$$<$$\ensuremath{p_T^\mathrm{assoc}}$$<$$2$~\ensuremath{\mathrm{GeV}/c}
      as a function of the number of participating nucleons (\ensuremath{N_{part}}) in PbPb collisions 
      at \ensuremath{\sqrt{s_{_{NN}}}} = 2.76~TeV.
      {\protect \cite{ref:HIN-10-002-PAS}} are also shown. 
      Statistical and systematic uncertainties are negligible. 
    }
    \label{fig:fig4}
  \end{center}
\vspace{-0.8cm}
\end{figure}

\begin{equation}
\label{eq:small_vn}
v_{\rm n}^{\rm f}(\ensuremath{p_T^\mathrm{trig}}) = \frac{C^{\rm f}_{\rm n}(\ensuremath{p_T^\mathrm{trig}},\ensuremath{p_T^\mathrm{assoc}})}{v_{\rm n}^{\rm f}(\ensuremath{p_T^\mathrm{assoc}})},
\end{equation}

\noindent where $v_{\rm n}^{\rm f}(\ensuremath{p_T^\mathrm{assoc}})$ can be calculated as
$\sqrt{C^{\rm f}_{\rm n}(\ensuremath{p_T^\mathrm{trig}},\ensuremath{p_T^\mathrm{assoc}})}$ 
for both \ensuremath{p_T^\mathrm{trig}} and \ensuremath{p_T^\mathrm{assoc}} 
in the same range as the one chosen for \ensuremath{p_T^\mathrm{assoc}}, which for this analysis
was selected as $1$$<$$\ensuremath{p_T^\mathrm{assoc}}$$<$$2$~\ensuremath{\mathrm{GeV}/c}.
The extracted $v_{\rm 2}^{\rm f}$ are shown in Fig.~\ref{fig:fig4}(a)
as a function of \ensuremath{p_T^\mathrm{trig}} 
for central and mid-peripheral collisions.
The results are found to be in good agreement with elliptic flow values measured by 
standard flow methods~\cite{ref:HIN-10-002-PAS}.
Fig.~\ref{fig:fig4}(b) gives the centrality dependence of the 
extracted flow coefficients at low $p_T$
for the higher harmonics $n$=2 to 5. The strong centrality dependence of $v_{\rm 2}^{\rm f}$ 
matches expectation from standard elliptic flow analyses, and the observed 
much weaker centrality dependence of the higher harmonics, as well as their 
observed relative strengths, awaits a more quantitative comparison to other data and theory.


\section*{References}
\begin{thebibliography}{10}

\bibitem{CMS:2008zzk}
  R.~Adolphi {\it et al.}  [CMS Collaboration],
  JINST {\bf 3}, S08004 (2008).

\bibitem{ref:HIN-11-001-PAS}
  CMS Collaboration, \texttt{arXiv:1105.2438}.

\bibitem{star:2005ja}
  STAR Collaboration, Phys. Rev. Lett. \textbf{ 95} (2005) 152301.

\bibitem{phenix:2007aa}
  PHENIX Collaboration, Phys. Rev. Lett. \textbf{ 98} (2007) 232302.
 
\bibitem{Alver:2008gk}
  PHOBOS Collaboration, Phys. Rev. \textbf{C81} (2010) 024904.

 \bibitem{star:2007pu}
  STAR Collaboration, J. Phys. G: Nucl. Part. Phys. \textbf{ 34} (2007) S679.
  
\bibitem{star:2010ridge}
  STAR Collaboration, Phys. Lett. \textbf{ B683} (2010) 123.

\bibitem{star:2009qa}
  STAR Collaboration, Phys. Rev.\textbf{ C80} (2009) 064912.

\bibitem{ref:HIN-10-002-PAS}
  CMS Collaboration, CMS Physics Analysis Summary, \textbf{ HIN-10-002} (2011).

\bibitem{ref:HIN-11-006-PAS}
  CMS Collaboration, CMS Physics Analysis Summary, \textbf{ HIN-11-006} (2011).

\bibitem{Alver:2009id}
  PHOBOS Collaboration, Phys. Rev. Lett. \textbf{ 104} (2010) 062301.

\bibitem{Alver:2010gr}
  B.~Alver and G.~Roland, Phys. Rev. \textbf{C81} (2010) 054905.

\bibitem{Voloshin:1994mz}
  S.~Voloshin and Y.~Zhang, Z. Phys. \textbf{ C70} (1996) 665.

\end{thebibliography}
\end{document}